\documentclass[%
 reprint,prl,
 amsmath,amssymb,longbibliography,aps,
]{revtex4-2}

\usepackage{natbib}
\usepackage{graphicx}
\usepackage{dcolumn}
\usepackage{bm}

\usepackage{graphicx}
\usepackage{dcolumn}
\usepackage{bm}
\usepackage{hyperref}
\usepackage{xcolor}
\usepackage{amsmath, amssymb, amsfonts}
\usepackage{mathtools}
\usepackage{subfigure}
\usepackage{mathrsfs}
\usepackage{comment}
\usepackage{amsthm}
\usepackage{ dsfont }

\newcommand{\bl}{\beta\lambda}
\newcommand{\eone}{{\epsilon_1}}
\newcommand{\eonep}{{\epsilon_1^\prime}}
\newcommand{\etwo}{{\epsilon_2}}
\newcommand{\etwop}{{\epsilon_2^\prime}}
\newcommand{\ethree}{{\epsilon_3}}

\newcommand{\ei}{{\epsilon_i}}
\newcommand{\eip}{{\epsilon_i^\prime}}

\newcommand{\rl}{{\rangle\langle}}
\newcommand{\beq}{\begin{equation}}
\newcommand{\eeq}{\end{equation}}
\newcommand{\beqn}{\begin{eqnarray}}
\newcommand{\eeqn}{\end{eqnarray}}
\newcommand{\p}{\partial_{\theta}}

\begin{document}

\title{
Thermodynamic Uncertainty of Work in Time-Dependently Driven Open Quantum Systems}

\author{Chulan Kwon}
\affiliation{Department of Physics, Myongji University, Yongin, Gyeonggi-Do,
17058,  Korea}
\email{ckwon58@gmail.com}

\date{\today}

\begin{abstract}
We derive a thermodynamic uncertainty relation for work in an open quantum system driven by a time-dependent protocol and subject to repeated projective energy measurements.  The protocol is represented by a sequence of protocol quenches separated by finite-time Lindblad evolution, so that work is accumulated at the quenches while dissipation occurs between them.  The resulting work statistics obey the Gallavotti-Cohen symmetry to give rise to the Crooks and Jarzynski fluctuation relations.  By perturbing the dissipative dynamics and combining the Cram\'er--Rao inequality with the quantum Fisher information, we obtain
$\mathrm{Var}\,W/[\tau\partial_\tau\langle W\rangle]^2\ge 1/F\ge\max(1/A,2/E)$, where $\langle W\rangle$ is the work expectation value, $\mathrm{Var}\,W$ is the variance of work, $F$ is the Fisher information, $A$ is the dynamical activity, $E$ is the dimensionless entropy production, and $\tau$ is the switching time interval between quenches.  The two thermodynamic bounds follow from two independent perturbations that generate the same response of the work statistics.  We illustrate the relation for a dissipative two-level system under square-wave and sinusoidal driving and show that the tighter thermodynamic bound can depend on the driving protocol and switching time scale.
\end{abstract}
\maketitle

Thermodynamic uncertainty relations (TURs) quantify a fundamental trade-off between the precision of fluctuating observables and the thermodynamic or kinetic resources required to sustain them.  For classical Markov processes, the original steady-state TURs bound current fluctuations in terms of entropy production and dynamical activity~\cite{barato2015,gingrich2016}.  Extensions to Langevin systems~\cite{dechant2018,hasegawa2019}, periodically driven systems~\cite{pietzonka2018,koyuk2019} and, subsequently, to arbitrary time-dependent driving from arbitrary initial states~\cite{koyuk_seifert2020,jslee2021,ckwon2022}. 
These results provide the classical starting point for uncertainty relations away from stationary conditions.

Quantum TURs have developed along a complementary route based on quantum estimation theory.  
The Fisher information (FI) quantifies the statistical distinguishability generated by a perturbation and is bounded by the quantum Fisher information (QFI)~\cite{wootters1981,braunstein_caves1994}. For Lindblad dynamics, the QFI can be evaluated using a pure-state system--reservoir construction~\cite{gammelmark_molmer2014}.
For Markovian open quantum systems, Hasegawa derived uncertainty relations for continuous measurements whose fluctuations are bounded by dynamical activity or entropy production~\cite{hasegawa2020}, and later formulated a TUR for general open quantum dynamics~\cite{hasegawa2021}.  Finite-time thermodynamic and kinetic bounds for quantum-jump observables were obtained by Vu and Saito~\cite{vu_saito2022,vu2025,kwon_unified2025}, while Nakajima and Utsumi clarified the role of the symmetric-logarithmic-derivative Fisher information in quantum kinetic uncertainty relations~\cite{nakajima2023}.  Related developments have addressed steady-state currents under coherent driving~\cite{menczel2021} and periodically operated quantum heat engines in the slow-driving regime~\cite{miller2021}.  Thus, quantum uncertainty relations now cover a broad range of monitored currents and trajectory observables as well as several classes of driven open systems.

The observable considered here is different: it is the work accumulated as an external protocol changes the system Hamiltonian.  Quantum work is conventionally characterized through projective energy measurements~\cite{talkner2007}, and work statistics and fluctuation relations in open quantum systems have been studied in a variety of settings.  Of particular relevance, Monnai established a TUR for the quantum work distribution of an externally perturbed harmonic oscillator coupled to reservoirs~\cite{monnai2022}.  Projective-energy-measurement statistics in a finite-time quantum heat engine have also been compared with known TUR bounds~\cite{mckeever2025}.  These results concern different models, driving protocols, and measurement settings.  In contrast, we consider a piecewise time-dependent Markovian open quantum system in which work is resolved directly from successive projective measurements of the system energy at protocol changes, while dissipation takes place during the intervening Lindblad evolution.  The measurements therefore determine both the stochastic work increments and the initial state for each subsequent dissipative interval.

We represent the time-dependent protocol by a sequence of piecewise-constant Hamiltonians.  Each protocol change is treated as an instantaneous quench, at which the corresponding energy difference is identified as work, while the evolution between quenches is generated by a time-independent Lindblad operator for the fixed Hamiltonian.  This construction provides a controlled discretization of a smooth protocol without requiring a continuously time-dependent dissipative generator~\cite{albash2012,kosloff2021}.  It also mimics an experimental situation in which a protocol device generates an external field with a finite switching time that cannot be made arbitrarily short.

Our central result combines the time-scale response characteristic of driven TURs with QFI bounds for open quantum dynamics.  We first construct the generating function of the accumulated work and establish the Gallavotti--Cohen symmetry, which yields the Crooks and Jarzynski fluctuation relations.  We then introduce an auxiliary perturbation parameter $\theta$ in the dissipative dynamics.  The Cram\'er--Rao inequality bounds the work fluctuations by the Fisher information (FI), while the quantum Fisher information (QFI) provides a measurement-independent upper bound on the FI.  Crucially, two independent perturbations generate the same response of the work statistics: both satisfy $\partial_\theta T_i=\tau\partial_\tau T_i$ for the finite-time transition probability $T_i$ in each dissipative interval.  For the first perturbation (R1), the QFI is the dynamical activity ${\cal A}$; for the second (R2), it is bounded by one half of the dimensionless entropy production ${\cal E}$.  Since the two perturbations produce the same work response, the two QFI bounds apply simultaneously and give
\begin{equation}
\frac{\mathrm{Var}\,{\cal W}}{[\tau\partial_\tau\langle{\cal W}\rangle]^2}
\ge \frac{1}{{\cal F}}
\ge \max\left(\frac{1}{{\cal A}},\frac{2}{{\cal E}}\right).
\label{QTUR_intro}
\end{equation}
This construction distinguishes the present result from uncertainty relations formulated primarily for continuously monitored currents or quantum-jump counting observables, and from work TURs restricted to specific solvable models or slow-driving regimes.  We illustrate the bound for a dissipative two-level system driven by square-wave and sinusoidal protocols, showing explicitly how the tighter of the activity and entropy-production bounds depends on the protocol and switching time scale.

We suppose a piecewise-constant protocol $\{\alpha_i\}_{i=0}^{N}$ with a switching interval $\tau$.  During each interval the Hamiltonian is fixed and the reduced dynamics is described by the corresponding Lindblad generator ${\cal L}_i$.  The interval $\tau$ is taken long enough for the Markovian and secular description to be applicable for the model under consideration, while neighboring protocol values are chosen sufficiently close when a smooth driving protocol is approximated.  The piecewise construction is therefore a controlled modeling choice, that is expected to be applicable for experiments using protocol devices with limited switching times. 

At each diabatic jump at time $t_i$, the change in $H$ is identified as work, whereas no heat is produced in the quenched system. Work is measured using a two-point projective measurement scheme, implemented by the consecutive application of two projectors,  $M(\epsilon_{i-1})=| \epsilon_{i-1} \rangle\langle \epsilon_{i-1}|$ for $t=t_i-0^+$ and $M(\epsilon_{i})=|\epsilon_{i} \rangle\langle \epsilon_{i}|$ for $t=t_i+0^+$, where $| \epsilon_{i} \rangle$ is an energy eigenstate of $H_i=H(\lambda_i)$. The combined measurement operator $M_i$ is given by a time-ordered product $M(\epsilon_i)M(\epsilon_{i-1})$ with a work increment $\epsilon_i-\epsilon_{i-1}$. For simplicity, we demonstrate the time evolution of the density operator through three measurement steps starting from the initial Boltzmann state associated with $H_0=H(\alpha_0)$. The density operator right after the third measurement at $t_3$ is given by $\rho(t_3)=M_3\left[ e^{\tau{\cal L}_2} M_2 \left[e^{\tau{\cal L}_1} M_1 [e^{\tau{\cal L}_0} e^{-\beta H_0}/Z_0]M_1^\dagger\right]M_2^\dagger\right]M_3^\dagger$ where ${\cal L}_i$ is the Lindblad generator for protocol $\alpha_i$ and $Z_0$ is the partition function at an inverse temperature $\beta$ associated with $H_0$. By inserting $e^{-\beta\lambda H_i}$ and $e^{-\beta\lambda H_{i-1}}$, respectively, in front of  $M(\ei)$ and $M(\epsilon_i)$ (not their adjoints) and taking the trace, we define the generating function 
\beq
{\cal G}_N(\lambda)=E\cdot G_N(\lambda)\cdot T_{N-1}\cdots G_{2}(\lambda)\cdot T_{1}\cdot G_1(\lambda)\cdot P_1
\label{matrix_form}
\eeq
where $G_i$, $T_i$ are $K\times K$ matrices for a $K$-level system and $E,~P_1$ are $K$-dimensional row and column vectors, respectively. The matrix elements are as follows: $G_i(\ei,\epsilon_{i-1};\lambda)=e^{-\beta\lambda(\ei-\epsilon_{i-1})}\left |\langle\epsilon_i|\epsilon_{i-1}\rangle \right|^2$ for work increment $w_i=\ei-\epsilon_{i-1}$ and $T_i(\eip,\ei)=\langle\epsilon_i^\prime | \left[e^{\tau{\cal L}_{i}}|\epsilon_i\rangle\langle\epsilon_i|\right]|\epsilon_i^\prime\rangle$. $T_i$ describes the finite-time transition probability for the evolution from $t_i$ to $t_{i+1}$.  The vector components are as follows: $E=(1,1,\cdots, 1)$ and $P_1(\epsilon_0)=\langle\epsilon_0 |e^{-\tau{\cal L}_0}e^{-\beta H_0} |\epsilon_0\rangle=e^{-\beta \epsilon_0}/Z_0$.  The cumulants are generated by $(-\partial_{\beta\lambda})^n \left. \ln {\cal G}_i(\lambda)\right |_{\lambda\to 0}$. In particular, this yields the expectation value of work for $n=1$ and the variance of work for $n=2$. The accumulated work $ {\cal W}$ along a measurement trajectory is given by ${\cal W}=\sum w_i$ for $w_i=\ei-\epsilon_{i-1}$. 
Writing $G_i^{(n)}=(-\partial_{\beta\lambda})^n G_i(\lambda)|_{\lambda\to 0}$, we find 
\beq
\langle {\cal W}\rangle=\sum_i^N E\cdot G_i ^{(1)}\cdot T_{i-1}\cdot P_{i-1}~
\label{expectation_work}
\eeq
where the recursion relation gives $P_{i-1}=G_{i-1}^{(0)}\cdot T_{i-2}\cdot P_{i-2}$ for $i\ge 3$ and  $P_1$ is given from Eq.~(\ref{matrix_form}). The variance of $\cal W$ 
is given similarly in terms of $G_i^{(n)}$ for $n=0,1,2$.

\begin{figure}[t]
    \centering
    \includegraphics[width=\columnwidth]{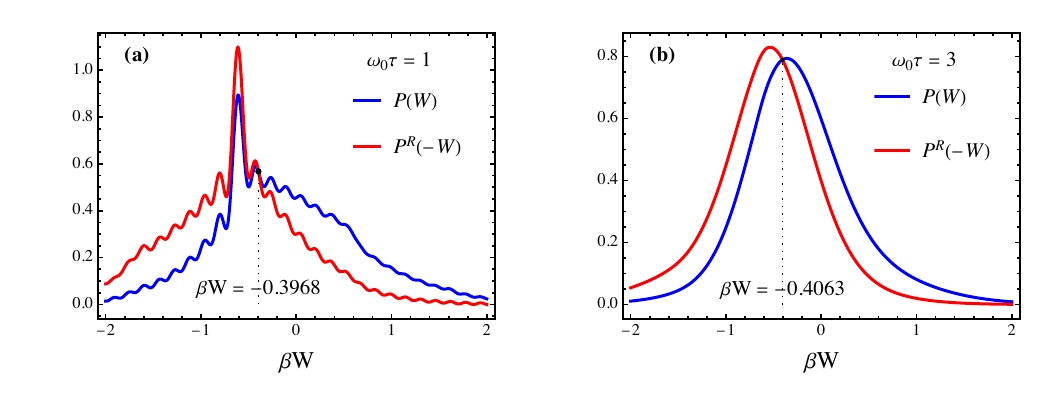}
    \caption{Forward work distribution $P(W)$ and reverse distribution
    $P^{\mathrm R}(-W)$ for $\alpha_i=2 \sin\left[2\pi i\tau/T\right]$. (a) $\omega_0\tau=1$, $\omega_0    T=100$, $N=125$ (b) $\omega_0\tau=3$, $\omega_0 T=300$, $N=125$. For both cases, $\beta\Delta F=-0.4063$, which is close to the indicated values of $\beta W$ at the intersections. }
    \label{fig:Crooks}
\end{figure}

We can show that ${\cal G}_N(\lambda)$ satisfies the Gallavotti-Cohen (GC) symmetry~\cite{gallavotti_cohen1995,kurchan1998,lebowitz_spohn1999} if the system is prepared in equilibrium associated with $H_0$:
\beq
{\cal G}_N(\lambda)={\cal G}^{\textrm{R}}_N(1-\lambda)e^{-\beta \Delta F}~,
\label{GC-symm}
\eeq
where $\Delta F=F_N-F_0$ is the difference of the free energies defined as $e^{-\beta F_{N,0}}=\textrm{Tr} e^{-\beta H_{N,0}}$. ${\cal G}^{\textrm R}_N(\lambda)$ is the generating function for the adjoint dynamics in which the protocol changes reversely in time as $\alpha_N\to\alpha_{N-1}\to\cdots\to\alpha_1\to\alpha_0$ and the time evolution during 
the $i$-th switching time interval is given by ${\cal L}^{\textrm{R}} _i={\cal L}^*_{N-i}$ where $*$ denotes complex conjugation. The proof is given in Sec. I, Supplementary Material (SM). The probability distribution of work fluctuation can be found from $P(W)=\beta/(2\pi)\int_{-\infty}^\infty d\lambda {\cal G}_N(i\lambda) e^{i\beta\lambda W}$. Similarly, $P^{\textrm R}(W)$ can be found from ${\cal G}^{\textrm R}_N(\lambda)$. As a consequence of the GC-symmetry, we get the Crooks detailed fluctuation theorem (FT)~\cite{crooks1999}
\beq
\frac{P(W)}{P^{\textrm R}(-W)} = e^{\beta(W-\Delta F)}~,
\label{crooks}
\eeq
and the Jarzynski integral FT~\cite{Jarzynski1997PRL,Jarzynski1997PRE}, ${\cal G}_N(1)=\langle e^{-\beta W}\rangle_i = e^{-\beta\Delta F}$ using ${\cal G}_N^{{\cal R}}(0)=1$.
Analogous fluctuation relations hold for classical stochastic systems~\cite{seifert2005,esposito2010}; here the distinction lies in the definition of the adjoint dynamics. Figure~\ref{fig:Crooks} compares the work distributions for the forward and reverse processes and illustrates the Crooks FT in Eq.~(\ref{crooks}).

From the perspective of estimation theory, we perturb the dynamics by a parameter $\theta$, which modifies the density operator as $\rho\to\rho_\theta$. For convenience, we introduce two parameters $\theta_1$ and $\theta_2$ and take the limit $\theta_1=\theta_2=\theta\to0$ at the end of the calculation. 
We follow the pure state formalism for the Lindblad dynamics proposed by Gammelmark and M{\o}lmer~\cite{gammelmark_molmer2014}. One can write the wave vector for the composite system and reservoir as
\beqn
|\psi_{\theta_1}(\tau)\rangle&=&\sum_{m_0,\ldots,m_{M-1}}V_{m_{M-1}}(\theta_1)\cdots V_{m_0}(\theta_1)
|\ei\rangle
\nonumber\\
&&\otimes ~| m_{M-1},\ldots,m_0\rangle~,
\nonumber\\
\langle\psi_{\theta_2}(\tau) |&=&\sum_{m_0,\ldots,m_{M-1}}\langle m_{M-1},\ldots,m_0|~
\nonumber\\
&&\otimes~\langle\ei |\tilde{V}^\dagger_{m_0}(\theta_2)\cdots \tilde{V}^\dagger_{m_{M-1}}(\theta_2)~,
\label{composite_wavevector}
\eeqn
where $m=0$ denotes no jump, $m>0$ labels a jump channel, and $|m_{M-1},\ldots,m_0\rangle$ records the corresponding reservoir trajectory at times $t_0,\ldots,t_{M-1}$. For $m>0$,
the left-acting operator is given by $V_m(\theta_1)=\sqrt{dt\,[1+h_m\theta_1]}L_m$ and the right-acting one by $\widetilde{V}_m^\dagger(\theta_2)=\sqrt{dt\,[1+h_m\theta_2]}L_m^\dagger$ where $L_m$ is the jump operator for transition channel $m$. For $m=0$ (no transition), $V_0(\theta_1)=I-idt H/\hbar-1/2\sum_{m>0}V_m^\dagger(\theta_1) V_m(\theta_1)$ and $\widetilde{V}_0^\dagger(\theta_2)=I+idt H/\hbar-1/2\sum_{m>0}\tilde{V}_m^\dagger(\theta_2) \tilde{V}_m(\theta_2)$ where $I$ is the identity operator. Taking the partial trace over the reservoir, we have $\textrm{Tr}_{\textrm{R}}|\psi_{\theta_1}(\tau)\rangle\langle\psi_{\theta_2}(\tau) |=\sum_{\{m_j\}}\!\!V_{m_{M\!-\!1}}\!(\theta_1)\cdots V_{m_0}\!(\theta_1)|\ei \rl \ei |\tilde{V}_{m_0}^\dagger\!(\theta_2)\cdots\tilde{V}_{m_{M\!-\!1}}^\dagger\!(\theta_2)$. In the $M\!\to\!\infty$ limit for $t_j\!=\!j(\tau/M)$, it leads to $\rho_\theta(\tau)=
e^{\tau {\cal L}_\theta}|\ei \rl \ei |$ where the perturbed generator is given as
\beqn
\lefteqn{{\cal L}_\theta(\rho)=-\frac{i}{\hbar}[H,\rho]}\nonumber\\
&&- \frac{1}{2}\!\!\sum_{m>0} \!\!
\left[(1\!+\!h_m\theta_1)L_m^\dagger L_m\rho+(1\!+\!h_m\theta_2)\rho L_m^\dagger L_m\right]
\nonumber\\
&&+\sum_{m>0}\sqrt{(1+h_m\theta_1)(1+h_m\theta_2)}L_m\rho L_m^\dagger~.
\label{generator_perturb}
\eeqn
At $\theta_1=\theta_2=0$, this reduces to the original Lindblad generator. 

We derive the Cram\'er--Rao inequality in Sec. II, SM:  
\beq
\frac{var {\cal W}}{ [\p\langle {\cal W}\rangle ]^2}\ge \frac{1}{\cal F}~.
\eeq
where we use the shorthand $\p f
=(\partial_{\theta_1}+\partial_{\theta_2})f |_{\theta_1,\theta_2 \to 0}$ is used. ${\cal F}$ is the Fisher information (FI) summed over the switching intervals. The quantum Fisher information (QFI), ${\cal QF}$, provides an upper bound on the FI. Both quantities are written as $\sum_{i=1}^{N-1} \sum_{\ei}f_i(\ei)P(\ei)$, where $f_i={\cal F}_i$ (${\cal QF}_i$) is a local production of FI (QFI). For an interval between $t_i$ and $t_{i+1}$, the local contributions follow from Eqs.~(S17) and (S28) in SM as
\beqn
{\cal F}_i(\ei)&=&\sum_{\eip}[\p T_i(\eip |\ei)]^2/T_i(\eip |\ei)~,
\label{FI_text}\\
{\cal QF}_i(\ei)&=&4\partial_{\theta_1}\partial_{\theta_2}\ln \textrm{Tr}~ e^{\tau {\cal L}_{i \theta}}|\ei \rl \ei|\Big |_{\theta_1=\theta_2=\theta\to 0}
\eeqn
where ${\cal L}_{i \theta}$ is given in Eq.~(\ref{generator_perturb}) with $H=H_i$. The local QFI, ${\cal QF}_i$, upper-bounds the corresponding local FI, ${\cal F}_i$, associated with any subsequent measurement at $t_{i+1}$~\cite{braunstein_caves1994}, as shown in Eq.~(S23), Sec. III, SM. In the present problem, the QFI can further be related to thermodynamic quantities such as the dynamical activity and entropy production. 

For a $K$-level system, let $|a\rangle$, $a=1,2,\ldots,K$, denote an energy eigenstate of the fixed Hamiltonian $H$ during a switching interval, with energy $\hbar\omega_a$. If channel $m$ denotes a transition from $a$ to $b$, we write $L_m=\sqrt{\gamma_{ba}}|b \rl a|$. For its conjugate transition from $b$ to $a$, we use the detailed balance condition $\gamma_{ab}=\gamma_{ba}e^{\beta\hbar(\omega_b-\omega_a)}$, which guarantees relaxation toward equilibrium as $\tau\to\infty$; for finite $\tau$, the state is generally nonequilibrium. Perturbation rules relating the QFI to dynamical activity and entropy production have been developed for steady states. We extend the two rules introduced by Hasegawa~\cite{hasegawa2020} to the present time-dependent states:
\begin{equation}
\textrm{R1}:~~h_{ba}=1~,~~
\textrm{R2}:~~h_{ba}=1-\sqrt{\frac{\gamma_{ab}\rho_{bb}}{\gamma_{ba}\rho_{aa}}}~,
\end{equation}
where the populations $\rho_{aa}=\rho_{aa}(t)$ evolve during each finite switching interval.  The two rules lead to different QFIs but, as shown below, to the same FI response relevant to the Cram\'er--Rao bound in Eq.~(\ref{QTUR_intro}). We also examine the rule (R3) by Vu~\cite{vu2025} in Secs.~IV and V, SM; it does not yield the same global bound after accumulation over the protocol intervals. 
We derive the key identity 
$\p T_i=\tau\partial_{\tau}T_i$ for both perturbation rules, shown in Eq.~(S37), Sec. IV, SM. Consequently, the left-hand side of the inequality and 
$1/{\cal F}$ are independent of the perturbation rule. We show in Sec. V, SM that the QFI for R1 equals the dynamical activity ${\cal A}$, whereas the QFI for R2 is bounded by ${\cal E}/2$, with ${\cal E}$ the dimensionless entropy production.  Since the FI is the same for the two perturbations, both upper bounds apply simultaneously.  Therefore, we find 
\beq
\frac{var {\cal W}}{ [\tau\partial_\tau\langle W\rangle ]^2}\ge \frac{1}{{\cal F}}\ge\textrm{max}
\left(
\frac{1}{{\cal A}},\frac{2}{{\cal E}}\right)~.
\label{QTUR}
\eeq

We illustrate our results for a two-level system with $H_i=H_0+H_i^\prime$. Let $| e \rangle$ and $| g\rangle$ be the excited and ground state of $H_0$. We consider $H_0=(\hbar\omega_0/2)(| e\rl e |-| g \rl g|)$ and $H_i^\prime=\hbar\omega_0\alpha_i(| e\rl g |+| g \rl e |)$. The energy eigenstates of $H_i$ are given as 
$| \epsilon_i^{+}\rangle=1/N_i (c_i |e\rangle+d_i |g \rangle)$, $| \epsilon_i^{-}\rangle=1/N_i(-d_i |e\rangle+c_i |g \rangle)$
 with $c_i=1+\sqrt{1+\alpha_i^2}$,  
 $d_i=\alpha_i$, and $N_i=\sqrt{c_i^2+d_i^2}$. The energy eigenvalues are $\epsilon_i^{\pm}=\pm \hbar\omega_i/2$ with $\omega_i=\omega_0\sqrt{1+\alpha_i^2}$. The jump operators are given by
\beqn
L_{i-}&=&\sqrt{\gamma_0 (1+N(\omega_i))}|\epsilon_i^{-}\rl \epsilon_i^{+}|~,
\nonumber\\
L_{i+}&=&\sqrt{\gamma_0 N(\omega_i)}|\epsilon_i^{+}\rl \epsilon_i^{-}|~,
\label{jump_operator_2level}
\eeqn
which follow from a standard optical-bath model for an open quantum system~\cite{breuer_petruccione2007}, where $N(\omega_i)=1/(e^{\beta\hbar\omega_i}-1)$ is the average number of photons in a mode with frequency $\omega_i$. Note that the detailed balance condition is met from $N(\omega_i)/(1+N(\omega_i))=e^{-\beta\hbar\omega_i}$. Using the vectorization $\rho\to \widehat{\rho}=\sum_{m,n}\rho_{mn}| m\rangle \otimes | n\rangle$, we obtain $e^{\tau \widehat{\cal L}_i}$ explicitly in Eq.~(S59), Sec.~VI, SM. We can then obtain the analytic expression for the generating function in Eq.~(\ref{matrix_form}) and derived quantities. 

\begin{figure}[t]
    \centering
    \includegraphics[width=\columnwidth]{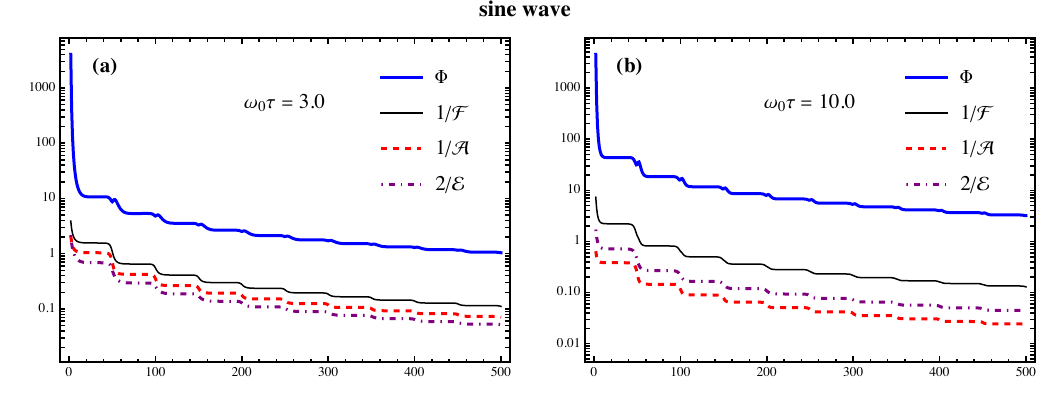}
    \caption{QTUR quantities for sinusoidal waves versus the measurement number $N$ for $\beta\hbar\omega_0=0.1,~\gamma_0/\omega_0=0.1$ and amplitude $A=10$.
    (a) $\omega_0T=300$ and
    $\omega_0\tau=3$. (b) 
    $\omega_0T=1000$ and $\omega_0\tau=10$.}
    \label{fig:QTUR_SineWave}
\end{figure}

For the numerical demonstrations, we work in the weak-coupling regime and set $\gamma_0/\omega_0=0.1$. For the parameter regime considered here, the switching interval is chosen so that $\omega_0\tau\gtrsim 1$, while remaining short on the time scale of the external protocol. For typical microscopic frequencies, $\omega_0^{-1}$ can lie in the range $10^{-15}$--$10^{-10}\,\mathrm{s}$, so this choice still permits $\tau$ to serve as a short time step in approximating a continuous protocol. We consider two periodic protocols with period $T$: a square wave and a sinusoidal wave. For a square wave, $\alpha_i=\alpha$ for odd $i$ and $0$ for even $i$. For a sinusoidal wave, we set $\alpha_i=A \sin [2\pi i\tau/T]$ with $\tau=T/M$ for a large integer $M$. 
\begin{figure}[t]
    \centering
    \includegraphics[width=\columnwidth]{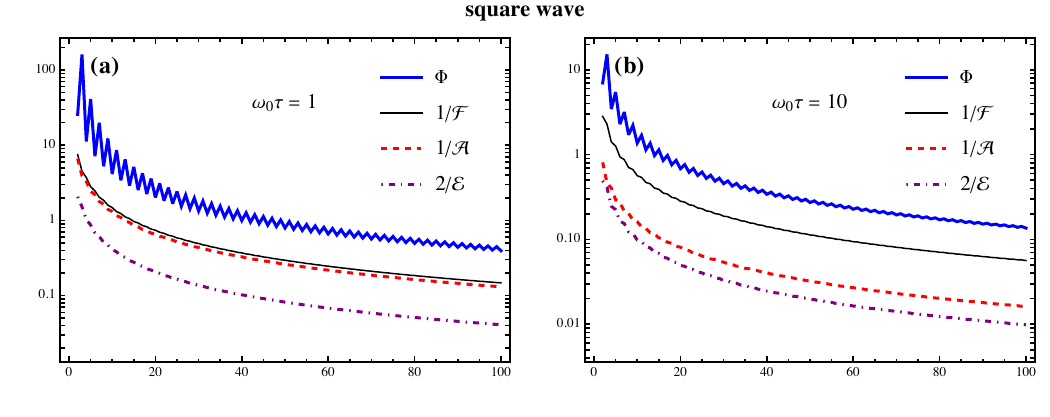}
    \caption{QTUR quantities for square wave protocols versus the measurement number $N$ for 
    $\beta\hbar\omega_0=1,~\gamma_0/\omega_0=0.1$ and amplitude $\alpha=10$. (a) $\omega_0T=2$ and $\omega_0\tau=1$; (b) $\omega_0T=20$ and $\omega_0\tau=10$.  The activity bound is tighter in both cases, $1/{\cal A}>2/{\cal E}$.}
    \label{fig:QTUR_SquareWave}
\end{figure}
 We present the curves of $\Phi=\mathrm{Var}\,{\cal W}/[\tau\partial_\tau \langle {\cal W}\rangle]^2$, $1/{\cal F}$, $1/{\cal A}$, and $2/{\cal E}$ in Fig.~\ref{fig:QTUR_SquareWave} for square waves and Fig.~\ref{fig:QTUR_SineWave} for sinusoidal waves, demonstrating the QTUR in Eq.~(\ref{QTUR}).  In Fig.~\ref{fig:QTUR_SquareWave},  $1/{\cal A}>2/{\cal E}$ in both panels, while the two thermodynamic bounds approach one another as $\tau$ increases.
In Fig.~ \ref{fig:QTUR_SineWave}, $1/{\cal A}>2/{\cal E}$ at the smaller switching interval, whereas $2/{\cal E}>1/{\cal A}$ at the larger one.  Thus the tighter thermodynamic bound switches from the activity bound to the entropy-production bound as the switching time scale is increased.

We have derived a finite-time thermodynamic uncertainty relation for accumulated work in a time-dependently driven open quantum system under repeated projective energy measurements.  Representing the protocol by quenches separated by finite-time Lindblad evolution allows work accumulation, dissipative relaxation, and measurement backaction to be incorporated within a single stochastic construction.  The resulting work statistics satisfy the GC symmetry yielding the Crooks and Jarzynski FTs.  The central result follows by applying two independent perturbations to the dissipative dynamics.  Both generate the same time-scale response, $\partial_\theta\langle W\rangle=\tau\partial_\tau\langle W\rangle$, while their QFIs are bounded separately by the dynamical activity and one half of the entropy production.  Combining these bounds yields Eq.~(\ref{QTUR}) without requiring a single perturbation whose QFI is globally $\min({\cal A},{\cal E}/2)$.  The two-level examples further show that the tighter thermodynamic bound can change with the driving protocol and switching time scale.  The result therefore extends QFI-based uncertainty relations to accumulated work resolved directly by projective measurements of the driven system.

\begin{acknowledgments}
This work was supported by the National Research Foundation of Korea (NRF) grant funded by the Korean government (MSIT) in 2025 (Grant No.~RS-2025-2106300413582110600002). OpenAI ChatGPT (GPT-5.6 Sol) was used to assist with literature organization, manuscript editing, and checks of mathematical and LaTeX consistency. The author directed the use of the tool and independently verified the scientific arguments, calculations, references, and final text.
\end{acknowledgments}

\paragraph{Data Availability Statement.} The Mathematica notebooks used for the numerical calculations and to generate the figures are publicly available from Ref.~\cite{KwonZenodo2026}.

\bibliographystyle{apsrev4-2}
\bibliography{bib}

\clearpage
\onecolumngrid
\setcounter{equation}{0}
\renewcommand{\theequation}{S\arabic{equation}}
\setcounter{section}{0}
\setcounter{secnumdepth}{1}
\renewcommand{\thesection}{\Roman{section}}

\begin{center}
{\large\bfseries Supplementary Material for\\[0.4em]
``Thermodynamic Uncertainty of Work in Time-Dependently Driven Open Quantum Systems''\par}
\vspace{1em}
{\normalsize Chulan Kwon\par}
\vspace{0.25em}
{\small Department of Physics, Myongji University, Yongin, Gyeonggi-Do 17058, Korea\par}
\end{center}
\vspace{1em}
We provide detailed derivations of important equations presented in the main text. While some of these derivations have appeared in previous works, we extend them to the present setting and include them here for completeness.  

\section{Gallavotti-Cohen symmetry}
\label{GC}

For simplicity of notation, we show the Gallavotti-Cohen symmetry for three-step measurement ($N=3$), which can be extended to an arbitrary $N$. We write the generating function 
\beqn
{\cal G}_3(\lambda)&=&\sum_{\textrm{all $\epsilon$'s}}\textrm{Tr} e^{-\bl H_3}M_\ethree e^{\bl H_2}M_\etwop \left[\left[e^{dt {\cal L}_{2}}\right]^n R_{2}(\lambda)\right ]M_\etwop M_\ethree
\nonumber\\
&=&\sum_{\textrm{all $\epsilon$'s}}\textrm{Tr} e^{\bl H_2}\left[(1+dt{\cal L}_2) A_2^{n-1}(\lambda)\right ]
 \underbrace{M_\etwop M_\ethree e^{-\bl H_3}M_\ethree M_\etwop}~,
\label{G3}
\eeqn
where the limit $n\to\infty$ is taken implicitly, so  $\left[e^{dt {\cal L}_{i-1}}\right]^n = e^{\tau {\cal L}_{i-1}}$, and we define $A_2^{n}(\lambda)=\left[e^{dt {\cal L}_2}\right]^nR_2(\lambda)$, and the summation is carried out over all possible projected eigenstates. We also use 
\beqn
R_2(\lambda)&=&\sum_{\textrm{all $\epsilon$'s}} e^{-\bl H_2}M_\etwo e^{\bl H_1}M_{\eonep}
 \left[\left[e^{dt {\cal L}_1}\right]^nR_1(\lambda)\right ]M_{\eonep}M_\etwo ~,
\nonumber\\
R_1(\lambda)&=&\sum_{\textrm{all $\epsilon$'s}} e^{-\bl H_1}M_{\eone} e^{\bl H_0}
 M_{\epsilon_0^\prime}\frac{e^{-\beta H_0}}{Z_0}M_{\epsilon_0^\prime}M_{\eone}~.
\eeqn

We introduce the adjoint dynamics where the protocol change is made reversely in time as $\alpha_N\to\alpha_{N-1}\to\cdots\to\alpha_1\to\alpha_0$.
We can start the time reversed process with
\beq
\tilde{R}_1(1-\lambda)=\sum_{\textrm{all $\epsilon$'s}e^{-\beta(1-\lambda)H_2} }\underbrace{M_{\etwop}e^{\beta(1-\lambda)H_3}M_\ethree\frac{e^{-\beta H_3}}{Z_3}M_\ethree M_{\etwop}}~.
\eeq
Comparing the terms in the underbraces in Eq.~(\ref{G3}) and in the above equation, we get the following:
\beq
{\cal G}_3(\lambda)=Z_3 \textrm{Tr}e^{\bl H_2}\left[(1+dt{\cal L}_2) A_2^{n-1}(\lambda)\right ]e^{\beta(1-\lambda)H_2} \tilde{R}_1(1-\lambda)~.
\eeq
We note
\beq
{\cal L}_2 A=-\frac{i}{\hbar}[H_2,A]-\frac{1}{2}\sum_{\omega_2}\{L(\omega_2)^\dagger L(\omega_2),A\}+\sum_{\omega_2}L(\omega_2)A L(\omega_2)^\dagger
\eeq
and use the property of jump operators:
\beq
e^{c\beta H} L(\omega) e^{-c\beta H}=e^{-c\beta \hbar\omega}L(\omega)~,~~
e^{c\beta H}L(\omega)^\dagger e^{-c\beta H}=e^{c\beta \hbar\omega}L(\omega)^\dagger~.
\eeq
Then, we get
\beqn
\textrm{Tr}e^{\bl H_2}{\cal L}_2 A e^{\beta(1-\lambda)H_2}R&=&\textrm{Tr} e^{\bl H_2}A e^{\beta(1-\lambda)H_2}\Big[-\frac{i}{\hbar}\left(RH_2-H_2 R\right)-\frac{1}{2}\sum_{\omega_2}\left(R L(\omega_2)^\dagger L(\omega_2)+L(\omega_2)^\dagger L(\omega_2)R\right)\nonumber\\
&&+\sum_{\omega_2>0} \left(e^{-\beta\hbar\omega_2}L^\dagger(\omega_2)R L(\omega 2)+e^{\beta\hbar\omega_2}L^\dagger(-\omega_2)R L(-\omega_2)\right)\Big]
\nonumber\\
&=&\textrm{Tr} e^{\bl H_2}A e^{\beta(1-\lambda)H_2}{\cal L}_2^*R
\eeqn
where $L(-\omega)=e^{-\beta\hbar |\omega |}L^\dagger( |\omega) |)$ is used. 
Therefore, we have
\beqn
{\cal G}_3(\lambda)
&=& Z_3\textrm{Tr}e^{\bl H_2}(1+dt {\cal L}_2)A_2^{n-2}(\lambda)e^{\beta(1-\lambda)H_2}
e^{dt {\cal L}_2^*}\tilde{R}(1-\lambda) 
\nonumber\\
&=&Z_3\textrm{Tr}e^{\bl H_2} R_2(\lambda)e^{\beta(1-\lambda)H_2}
(e^{dt {\cal L}_2^*})^n\tilde{R}(1-\lambda)
\nonumber\\
&=&Z_3 \sum_{\textrm{all $\epsilon$'s}}\textrm{Tr} M_{\etwo} e^{\bl H_1} M_{\eonep} \left[(e^{dt {\cal L}_1})^n R_1(\lambda)\right]
M_{\eonep} M_{\etwo} e^{\beta(1-\lambda)H_2}(e^{dt {\cal L}_2^*})^n\tilde{R}(1-\lambda)
\nonumber\\
&=&Z_3 \sum_{\textrm{all $\epsilon$'s}}\textrm{Tr} e^{\bl H_1} A_1^{n}(\lambda) e^{\beta(1-\lambda)H_1}
 e^{-\beta(1-\lambda)H_1}M_{\eonep}e^{\beta(1-\lambda)H_2}M_{\etwo} 
(e^{dt {\cal L}_2^*})^n\tilde{R}_1(1-\lambda)M_{\etwo}M_{\eonep}
\nonumber\\
&=&Z_3 \textrm{Tr} e^{\bl H_1} A_1^{n}(\lambda) e^{\beta(1-\lambda)H_1}\tilde{R}_2(1-\lambda)~,
\eeqn
where the definition for the next step for the reverse process is found as
\beq
\tilde{R}_2(1-\lambda)=\sum_{\textrm{all $\epsilon$'s}}e^{-\beta(1-\lambda) H_1}M_{\eonep}e^{\beta(1-\lambda) H_2}M_{\etwo}(e^{dt {\cal L}_2^*})^n\tilde{R}_1(1-\lambda)M_{\etwo}M_{\eonep}~.
\eeq
Repeating the above process, we get
\beqn
{\cal G}_3(\lambda)&=&Z_3\sum_{\textrm{all $\epsilon$'s}}\textrm{Tr} e^{\bl H_1}R_1(\lambda)e^{\beta(1-\lambda)H_1}(e^{dt{\cal L}_1^*})^n\tilde{R}_2(1-\lambda)\nonumber\\
&=&Z_3\sum_{\textrm{all $\epsilon$'s}}\textrm{Tr} M_{\eone} e^{\bl H_0} M_{\epsilon_0^\prime}\frac{e^{-\beta H_0}}{Z_0}M_{\epsilon_0^\prime}M_{\eone}e^{\beta(1-\lambda)H_1} (e^{dt{\cal L}_1^*})^n\tilde{R}_2(1-\lambda)
\nonumber\\
&=&\frac{Z_3}{Z_0}\sum_{\textrm{all $\epsilon$'s}}\textrm{Tr}e^{\beta(1-\lambda)H_0}M_{\epsilon_0^\prime}e^{-\beta(1-\lambda)H_1}M_{\eone}
 (e^{dt{\cal L}_1^*})^n\tilde{R}_2(1-\lambda)M_{\eone} M_{\epsilon_0^\prime}
\nonumber\\
&=&\frac{Z_3}{Z_0}\textrm{Tr}\tilde{R}_3(1-\lambda)=\frac{Z_3}{Z_0}{\cal G}^{\textrm{R}}_3(1-\lambda)~,
\eeqn
which is the proof of the GC-symmetry. 

\section{Cram\'er--Rao inequality}
\label{CR-inequality}
Let $\Delta {\cal W}$ be equal to $\sum_i^N \Delta w_i$ where $\Delta w_i=w_i-\langle w_i\rangle_\theta$.  For simplicity, we consider $N=3$. Eq.~(2) in the main text is modified for perturbation as 
\beq
\langle {\cal W}\rangle_\theta =\sum_{i=1}^{N}E\cdot G_i^{(1)}\cdot T_{\theta,i-1}
\cdot P_{\theta, i-1}
\eeq
where $G_i^{(0)}(\ei,\epsilon_{i-1}^\prime)=|\langle \ei |\epsilon_{i-1}^\prime \rangle |^2=p(\ei,\epsilon_{i-1}^\prime)$ and $G_i^{(1)}(\ei,\epsilon_{i-1}^\prime)=(\ei-\epsilon_{i-1}^\prime)p(\ei,\epsilon_{i-1}^\prime)$.
Note that $E\cdot G_N^{(0)}\cdot T_{\theta,N-1}\cdots G_{i+1}^{(0)}\cdot T_{\theta,i}=E$ and
 $P_{\theta, i-1}=W_{i-1}^{(0)}\cdot T_{\theta,i-2}\cdots W_{2}^{(0)}\cdot T_{\theta,1}\cdot P_1$. Then,
\beqn
\langle {\cal W}\rangle_\theta
&=&\sum_{i=1}^N E\cdot G_N^{(0)}\cdot T_{\theta,N-1}\cdot G_{N-1}^{(0)}\cdot T_{\theta,N-2}\cdots 
G_{i+1}^{(0)}\cdot T_{\theta,i}\cdot G_i^{(1)}\cdot T_{\theta, i-1}\cdot  G_{i-1}^{(0)}\cdot T_{\theta,i-2}
\cdots G_2^{(0)} \cdot T_{\theta,1}\cdot P_1
\nonumber\\
&=&\sum_{\textrm{all $\epsilon$'s}}\! \! \left[\sum_{i=0}^3\! (\epsilon_i-\epsilon_{i-1}^\prime)\right] p(\epsilon_3 ,\epsilon_2^\prime) T_\theta(\epsilon_2^\prime,\epsilon_2)
p(\epsilon_2 ,\epsilon_1^\prime)T_\theta(\epsilon_1^\prime,\epsilon_1)p(\epsilon_1,\epsilon_0^\prime)P_1(\epsilon_0^\prime)
\nonumber\\
&=& \sum_{\textrm{all $\epsilon$'s}} \left(\sum_{i=0}^3 w_i \right)
P_\theta(\epsilon_3,\epsilon_{2}^\prime,\epsilon_{2},\epsilon_1^\prime, \epsilon_1,\epsilon_0^\prime)
\eeqn
where $w_i=\ei-\epsilon_{i-1}^\prime$  and 
\beq
P_\theta(\epsilon_3,\epsilon_{2}^\prime,\epsilon_{2},\epsilon_1^\prime, \epsilon_1,\epsilon_0^\prime)=p(\epsilon_3 ,\epsilon_2^\prime) T_\theta(\epsilon_2^\prime,\epsilon_2)
p(\epsilon_2 ,\epsilon_1^\prime)T_\theta(\epsilon_1^\prime,\epsilon_1)p(\epsilon_1,\epsilon_0^\prime)P_1(\epsilon_0^\prime)
\eeq

Now, we have
\beqn
sum_{\textrm{all $\epsilon$'s}}\left(\sum_{i=0}^3\Delta w_i\right) \p P_\theta(\epsilon_3,\epsilon_{2}^\prime,\epsilon_{2},\epsilon_1^\prime, \epsilon_1,\epsilon_0^\prime)
&=&-\sum_{\ei,\epsilon_{i-1}^\prime}\!\! \left(\sum_{i=0}^3 \p\Delta w_i\right) P_\theta(\epsilon_3,\epsilon_{2}^\prime,\epsilon_{2},\epsilon_1^\prime, \epsilon_1,\epsilon_0^\prime)
\nonumber\\
&=&\p \langle {\cal W}\rangle_\theta~.
\eeqn
Using $\p \left[E\cdot G_N^{(0)}\cdot T_{\theta,N-1}\cdots G_{i+1}^{(0)}\cdot T_{\theta,i}\right]=0$, we find
\beqn
\left[\left.\p \langle {\cal W}\rangle_\theta\right |_{\theta\to 0}\right]^2
&=&\left[\sum_{\textrm{all $\epsilon$'s}}\left(\sum_{i=1}^3 \Delta w_i\right)\left(\sum_{i=1}^{2}\p\ln T(\eip,\ei)\right) P_{\theta=0}(\epsilon_3,\epsilon_{2}^\prime,\epsilon_{2},\epsilon_1^\prime, \epsilon_1,\epsilon_0^\prime)\right]^2
\nonumber\\
&\le&\underbrace{\left\langle\left(\sum_{i=1}^3\Delta w_i\right)^2\right\rangle}_{var {\cal W}}\times\underbrace{ \left\langle\left(\sum_{i=1}^2 \p\ln T(\eip,\ei)\right)^2\right\rangle}_{{\cal F}}
\eeqn
where $\p \ln T=
\p\ln T_\theta |_{\theta\to 0}$ and $\langle \cdots\rangle=\sum_{\textrm{all $\epsilon$'s}}(\cdots)P(\epsilon_3,\etwop,\etwo,\eonep,\eone,\epsilon_0^\prime)$. The inequality comes from the Cauchy-Schwarz inequality: $\left|\sum_i p_iA_i B_i\right |^2\le\sum_i p_i|A_i|^2\cdot\sum_i p_i|B_i|^2$.
$var W$ denotes the variance of work increment and ${\cal F}$ the Fisher information. ${\cal F}$ is more clarified as
 \beqn
 {\cal F}&=&\sum_{\etwop,\etwo}\frac{\left(\p T(\etwop |\etwo)\right)^2} {T(\etwop |\etwo)} P(\etwo)+\sum_{\eonep,\eone}\frac{\left(\p T(\eonep |\eone)\right)^2} {T(\eonep |\eone)} P(\eone)
 \nonumber\\
 &&+2\sum_{\etwo,\eonep}\!\!\underbrace{\left[\!\sum_{\etwop} \p T(\etwop|\etwo)\!\right]}_{0}\!\!p_2(\etwo|\eonep) \p T(\eonep|\eone)P(\eone)
 \eeqn
where $P(\etwo)=\sum_{\textrm{$\epsilon$'s}}^\prime P(\etwo,\eonep,\eone,\epsilon_0^\prime)$ for $\sum_{\textrm{$\epsilon$'s}}^\prime$ denoting the summation over all $\epsilon$'s except for $\etwo$, and similarly $P(\eone)=\sum_{\epsilon_0^\prime}p(\eone,\epsilon_0^\prime)$. For arbitrary $N$, we have
\beq
{\cal F}=\sum_{i=1}^{N-1}\sum_{\epsilon_i^\prime}\frac{\left(\p T(\epsilon_i^\prime |\epsilon_i)\right)^2} {T(\epsilon_i^\prime |\epsilon_i)} P(\epsilon_i)~,
\label{FI_app}
\eeq
where $P(\ei)$ is the $\ei$-component of $P_i$, which is recursively found as $G_i^{(0)}\cdot T_{i-1} \cdot P_{i-1}$ given $P_1$
which is the total Fisher information produced during each switching time interval. Then, we have the Cram\'er--Rao inequality
\beq
\frac{var {\cal W}}{[\p\langle {\cal W}\rangle]^2}\ge\frac{1}{{\cal F}}
\label{inequality}
\eeq         
where $\p \langle W\rangle=\lim_{\theta\to 0}\p \langle W\rangle_\theta$.

\section{Quantum Fisher information}
\label{QFI_section}
The finite-time transition probability (propagator) between projected states in the $i$-th interval is given as 
\beq
T_\theta(\epsilon_i^\prime|\epsilon_i)= \langle\eip |\left[e^{\tau {\cal L}_{\theta i}} |\ei \rl\ei |\right] |\eip\rangle=\textrm{Tr} M_{\eip}\rho_{\theta,\ei}
\eeq
where $\rho_{\theta,\ei}=e^{\tau {\cal L}_{\theta i}} |\ei \rl\ei |$. Then, the Fisher information produced during the interval given $\ei$ is given by 
\beqn
{\cal F}_i&=&\sum_{\eip}\frac{[\textrm{Tr} M_{\eip}\p\rho_{\ei}]^2}{\textrm{Tr} M_{\eip}\rho_{\theta,\ei}}~.
\eeqn

Following the work by Braunstein and Caves~\cite{braunstein_caves1994}, we derive the quantum Fisher information (QFI).
Using the symmetric logarithmic derivative (SLD),
\beq
\p \rho=\frac{1}{2}\left[\rho L(\p\rho)+L(\p\rho)\rho\right]~,
\eeq
we find
\beqn
[\textrm{Tr} M_{\eip}\p\rho_{\ei}]&=&\frac{1}{2}\textrm{Tr}\left(M_{\eip}\rho_{\ei}L+M_{\eip}L\rho_{\ei}\right)=\frac{1}{2}\textrm{Tr}\left(\rho_{\ei}LM_{\eip}+(\rho_{\ei}LM_{\eip})^\dagger\right)
\nonumber\\
&=&\textrm{Re} \textrm{Tr}\rho_{\ei}LM_{\eip}~.
\eeqn
Then, we find
\beqn
{\cal F}_i&\le&\sum_{\eip}\frac{\left|\textrm{Tr}\rho_{\ei}LM_{\eip}\right|^2}{\textrm{Tr} M_{\eip}\rho_{\ei}}
\nonumber\\
&=&\sum_{\eip}\frac{\left|\textrm{Tr}\rho_{\ei}^{1/2}LM_{\eip}^{1/2}\cdot M_{\eip}^{1/2}\rho_{\ei}^{1/2}\right|^2}{\textrm{Tr} M_{\eip}\rho_{\ei}}
\nonumber\\
&\le&\sum_{\eip}\textrm{Tr}\rho_{\ei}LM_{\eip}L
\nonumber\\
&=&\textrm{Tr}L\rho_{\ei}L={\cal QF}_i~,
\label{QFI_local}
\eeqn
where $|\textrm{Tr} AB|^2\le\textrm{Tr}AA^\dagger \cdot \textrm{Tr}BB^\dagger$ and  $\sum_{\eip}M_{\eip}=I~(\textrm{identity})$ are used. 

This expression of QFI holds both for mixed and pure states since it is driven without specifying ${\cal L}_i$ for $\rho_{\theta,\ei}=e^{\tau {\cal L}_{\theta i}} |\ei \rl\ei |$. For a mixed state, it is nontrivial to handle the SLD. For a pure state, we have a well-known way to find the QFI. We consider the perturbation of a state retaining normalization: $\langle \psi+d\psi |\psi +d\psi\rangle=1$. Then, $\langle d\psi |\psi\rangle+ \langle \psi | d\psi\rangle=-\langle d\psi | d\psi\rangle$ is of higher order, so neglected in the following derivation. Introducing vertical component $| d\psi_{\perp}\rangle=| d\psi\rangle-|\psi\rl \psi | d\psi\rangle$, we get
\beqn
d\rho&=&|\psi \rl d \psi |+| d\psi \rl  \psi |
\nonumber\\
&=& |\psi \rl d \psi_{\perp}|+| d\psi_{\perp} \rl  \psi | +\textrm{higher order}~.
\eeqn 
It can be shown that $ d\rho=L(d\rho)/2$,
\beq
L(d\rho)=2(|\psi \rl d \psi_{\perp}|+| d\psi_{\perp}      \rl  \psi | )~,
\eeq
which is shown to satisfy $d\rho=(\rho L+L\rho)/2$ by using $\langle\psi_{\perp}|\psi\rangle=0$. Then, the quantum Fisher information can be found as
\beqn
\cal QF
&=&\textrm{Tr}L(\p \rho)\rho L(\p\rho)
\nonumber\\
&=&\frac{1}{2}\textrm{Tr}(L(\p\rho)\rho L(\p\rho)+L(\p\rho)L(\p\rho)\rho)
\nonumber\\
&=&\textrm{Tr}L(\p\rho)\p\rho\nonumber\\
&=&2\textrm{Tr}\left(|\psi \rl \p \psi_{\perp}\!|+| \p\psi_{\perp} \rl  \psi |\right)
\left(|\psi \rl \p \psi_{\perp}|+| \p\psi_{\perp} \rl  \psi |\right)
\nonumber\\
&=&4\langle \p\psi_{\perp} |\p\psi_{\perp}\rangle
\nonumber\\
&=& 4\Big[\langle\p\psi | \p\psi\rangle-\langle \p \psi |\psi\rl \psi | \p\psi\rangle\Big]
\eeqn
where it is used that $\textrm{Tr} | \p\psi_{\perp} \rl \p \psi_{\perp}|=\langle\p \psi_{\perp}| \p\psi_{\perp}\rangle$.

Using Eq.~(6) in the main text and 
\beq
\langle\psi_{\theta_2}(\tau)|\psi_{\theta_1}(\tau)\rangle=\textrm{Tr}e^{\tau{\cal L}_i(\theta_1,\theta_2)}|\ei \rl \ei\rangle=T_{\theta_1,\theta_2}(\ei)~,
\eeq
we find
\beqn
{\cal QF}_i(\ei)&=&4\Big[ \partial_{\theta_1}\partial_{\theta_2}T_{\theta_1,\theta_2}(\ei)
-\partial_{\theta_1}T_{\theta_1,\theta_2}(\ei)\cdot \partial_{\theta_2}T_{\theta_1,\theta_2}(\ei)\Big]\Big|_{\theta_1,\theta_2\to 0}
\nonumber\\
&=&4~\partial_{\theta_1}\partial_{\theta_2}\ln ~T_{\theta_1,\theta_2}(\ei)\Big|_{\theta_1,\theta_2\to 0}
\label{QFI_formula}
\eeqn
Then, the total QFI is given as
\beq
{\cal QF}=\sum_{i=1}^{N-1}{\cal QF}_i P(\ei)~.
\label{QFI_app}
\eeq

\section{Properties of the first derivative}
\label{first_derv}

The density operator (DO) at the beginning of the switching time interval is diagonal in energy basis since the system is collapsed to a energy eigenstate by measurement. 
We consider  $L_{ba}=\sqrt{\gamma_{ba}}~ |b\rl a|$ and $H=\sum_a\hbar\omega_a |a \rl a |$ where the Einstein convention is used that the repeated indices are to be summed.  We investigate the subsequent evolution of the density operator (DO). From the perturbed generator in Eq.~(7) in the main text for $\theta_1=\theta_2=0$, using the Einstein convention for summation, we get
\beqn
\left[{\cal L}\rho\right]_{ab}&=&-\frac{1}{2}\gamma_{cd}\left[\langle a | d \rl c | c\rl d | e \rangle 
\rho_{eb}+\rho_{ae}\langle e|  d \rl c |c \rl d | b\rangle \right]
+\gamma_{cd}\langle a | c \rl d | d' \rangle\rho_{d' e}\langle e | d\rl c | b\rangle
\nonumber\\
&=& -\delta_{ab} \rho_{bb}\sum_c\gamma_{ca}+\delta_{ab}\sum_c\gamma_{ac}\rho_{cc}
\label{diagonal}
\eeqn 
Therefore, the DO remains diagonal as the initial DO is diagonal. 

We need to calculate 
\beqn
\p T(\epsilon^\prime |\epsilon)&=&\lim_{\theta\to 0}\p \textrm{Tr}M_{\epsilon^\prime}\left[e^{\tau{\cal L}_\theta} M_{\epsilon}\right]
\nonumber\\
&=&\int_0^\tau dt~ \textrm{Tr}M_{\epsilon^\prime}e^{(\tau-t){\cal L}} \p{\cal L}\underbrace{ \left[e^{t{\cal L}}M_{\epsilon}\right]}_{\rho(t)}
\eeqn
Note that $\p {\cal L}(\theta_1,\theta_2)=\partial_{\theta_1}{\cal L}\frac{d\theta_1}{d\theta} +\partial_{\theta_2}{\cal L}\frac{d\theta_2}{d\theta}\xrightarrow{\theta_1=\theta_2=\theta\to 0}=\p {\cal L}(\theta,\theta)|_{\theta\to 0}$. 
Eq.~(7) in the main text can be rewritten
\begin{eqnarray}
{\cal L}_\theta(\theta_1,\theta_2)\rho
&=&-\frac{i}{\hbar}[H,\rho]- \frac{1}{2}\sum_{b,a} 
\left[(1+h_{ba}\theta_1)L_{ba}^\dagger L_{ba}\rho+(1+h_{ba}\theta_2)\rho L_{ba}^\dagger L_{ba}\right]
\nonumber\\
&&+\sum_{b,a}\sqrt{(1+h_{ba}\theta_1)(1+h_{ba}\theta_2)}L_{ba}\rho L_{ba}^\dagger~.
\label{generator_perturbed_app}
\end{eqnarray}
Plugging $\rho=\sum_a \rho_{aa}| a \rl a|$ from Eq.(\ref{diagonal}), we have 
\beqn
\p {\cal L}&=& -\sum_{b,a}\gamma_{ba}h_{ba}(\rho_{aa} |a\rl a|-\rho_{aa} | b \rl b |)
\nonumber\\
&=&-\sum_{b,a}|a\rl a|(\gamma_{ba}h_{ba}\rho_{aa}-\gamma_{ab}h_{ab}\rho_{bb})
\label{deriv_generator}
\eeqn
Using the final line in Eq.~(\ref{diagonal}), we find
\beq
\p{\cal L}\rho-{\cal L}\rho=
-\sum_{b,a}|a\rl a|
\left[\gamma_{ba}(h_{ba}-1)\rho_{aa}-\gamma_{ab}(h_{ab}-1)\rho_{bb}\right]~.
\eeq
We examine the perturbation rules proposed by Hasegawa~\cite{hasegawa2020}.
For R1: $h_{ba}=1$, it vanishes. For R2: $h_{ba}=1-\sqrt{(\gamma_{ab}\rho_{bb})/(\gamma_{ba}\rho_{aa})}$
\beqn
\p{\cal L}\rho-{\cal L}\rho&=&\sum_{b,a}|a\rl a|\left[ \gamma_{ba}\sqrt{\frac{\gamma_{ab}\rho_{bb}}{\gamma_{ba}\rho_{aa}}}\rho_{aa}-\gamma_{ab}\sqrt{\frac{\gamma_{ba}\rho_{aa}}{\gamma_{ab}\rho_{bb}}}\rho_{bb}\right]
\nonumber\\
&=& 0~.
\eeqn
We also examine the rule R3 considered by Vu~\cite{vu2025}: 
\[h_{ba}=\frac{\gamma_{ba}\rho_{aa}-\gamma_{ab} \rho_{bb}}{\gamma_{ba}\rho_{aa}+\gamma_{ab} \rho_{bb}}\]
\beq
\p{\cal L}\rho-{\cal L}\rho
=2\sum_{b,a}| a \rl a |
\left[
\frac{\gamma_{ba}(\gamma_{ab}\rho_{bb})\rho_{aa}-\gamma_{ab}(\gamma_{ba}\rho_{aa})\rho_{bb}}
{\gamma_{ba}\rho_{aa}+\gamma_{ab}\rho_{aa}}
\right]
=0
\eeq

Therefore, we find an important result
\beqn
\p T(\epsilon^\prime|\epsilon)&=&\int_0^\tau dt~ \textrm{Tr}M_{\epsilon^\prime}e^{(\tau-t){\cal L}}{\cal L}[e^{t{\cal L}}M_\epsilon] 
\nonumber\\
&=&\lim_{\theta\to 0}\textrm{Tr}M_{\epsilon^\prime}\p[e^{\tau(1+\theta){\cal L}}M_\epsilon] 
\nonumber\\
&=&\tau\partial_\tau T(\epsilon^\prime|\epsilon)~.
\label{first_deriv_app}
\eeqn
 
\section{Quantities for the QTUR}
\label{Qtur_quantities}

Using the results from the previous sections, we find the quantities for QTUR. First, we have
\beqn
\p\langle W\rangle&=&\tau\partial_\tau\langle W\rangle~,
\label{work_deriv}\\
{\cal F}&=&\sum_{i=1}^{N-1}\sum_{\eip,\ei}\frac{\left[\tau\partial_\tau T(\eip |\ei)\right]^2}{ T(\eip |\ei)}P(\ei)
\label{FI_tau}
\eeqn

We obtain a more explicit expression for the local QFI in Eq.~(\ref{QFI_formula}) in the time evolution for interval $\tau$. 
For the QFI in Eq.~(\ref{QFI_formula}), we first need to calculate
\beqn
\partial_{\theta_1}\!\textrm{Tr}e^{\tau{\cal L}_\theta}\Big|_{\theta_1=\theta_2=\theta\to 0}
&=&\int_0^\tau dt\textrm{Tr} e^{(\tau-t){\cal L}}\left[(\partial_{\theta_1}{\cal L}) e^{t{\cal L}}M_{\epsilon}\right]
\nonumber\\
&=&\int_0^\tau dt\textrm{Tr} \left(\partial_{\theta_1}{\cal L}\right)\rho(t)
\nonumber\\
&=&\int_0^\tau dt\sum_{b,a}h_{ba}\textrm{Tr}\left[-\frac{1}{2} L_{ba}^\dagger L_{ba}\rho+\frac{1}{2}L_{ba}\rho L_{ba}^\dagger \right]~.
\nonumber\\
&=&0~. 
\eeqn
Similarly, $\partial_{\theta_2}\!\textrm{Tr}e^{\tau{\cal L}_\theta}\Big|_{\theta_1=\theta_2=\theta\to 0}=0$. Second, we need to calculate
\beqn
\partial_{\theta_1} \partial_{\theta_2}\!\textrm{Tr}e^{\tau{\cal L}_\theta}\Big|_{\theta_1=\theta_2=\theta\to 0}
&=&\textrm{Tr}\!\!\int_0^\tau \!\!\!dt\!\!\!  \int_0^t \!\!\!ds~ e^{(\tau-t){\cal L}} \left(\partial_{\theta_1}{\cal L}\right) e^{(t-s){\cal L}} \left(\partial_{\theta_2}{\cal L}\right) e^{s{\cal L}}M_\epsilon 
\nonumber\\
&&+\textrm{Tr}\!\!\int_0^\tau \!\!\!dt\!\!\!  \int_0^t \!\!\!ds~ e^{(\tau-t){\cal L}} \left(\partial_{\theta_2}{\cal L}\right) e^{(t-s){\cal L}} \left(\partial_{\theta_1}{\cal L}\right) e^{s{\cal L}}M_\epsilon 
+\textrm{Tr}\!\!\int_0^\tau \!\!\!dt\!\!~ e^{(\tau-t){\cal L}}\left(\partial_{\theta_1}\!\partial_{\theta_2}{\cal L}\right)e^{t{\cal L}}M_\epsilon
\nonumber\\
&=&\frac{1}{2}\textrm{Tr}\!\!\int_0^\tau \!\!\!dt\!\!\!  \int_0^t \!\!\!ds~ e^{(\tau-t){\cal L}} \left(\partial_{\theta}{\cal L}\right) e^{(t-s){\cal L}} \left(\partial_{\theta}{\cal L}\right) e^{s{\cal L}} M_\epsilon
+\textrm{Tr}\!\!\int_0^\tau \!\!\!dt\!\!~ e^{(\tau-t){\cal L}}\left(\partial_{\theta_1}\!\partial_{\theta_2}{\cal L}\right)e^{t{\cal L}}M_\epsilon~.
\eeqn
where $\theta_1=\theta_2=\theta$ is implicitly given, so $\partial_{\theta_1}=\partial_{\theta_2}=1/2\partial_\theta$. Note that $\partial_\theta^2{\cal L}=0$, but $\partial_{\theta_1}\!\partial_{\theta_2}{\cal L}\neq 0$. The first term in the last line is equal to $(1/2)\partial_\theta^2 \textrm{Tr}\rho_\theta(\tau)=0$.
Using the expression for the perturbed generator shown in Eq.~(7) in the main text, we have
\beqn
{\cal QF}_i&=&4 \textrm{Tr}\!\!\int_0^\tau \!\!\!dt\!\!~ e^{(\tau-t){\cal L}}\left(\partial_{\theta_1}\!\partial_{\theta_2}{\cal L}\right)e^{t{\cal L}}M_{\epsilon_i}
\nonumber\\
&=&\textrm{Tr}\int_0^\tau dt ~ \sum_{b,a} h_{ba}^2 L_{ba} \rho(t) L_{ba}^\dagger
\nonumber\\
&=&\int_0^\tau dt ~ \sum_{b,a}h_{ba}^2\gamma_{ba}\rho_{aa}(t)
\label{QFI_detail}
\eeqn
where $\rho(t)=e^{t{\cal L}}M_{\epsilon_i}$ for a starting collapsed state $|\ei\rangle$. 

For R1 with $h_{ba}=1$,  Eq.~(\ref{QFI_detail}) leads to
\beq
{\cal QF}_i=\int_0^\tau dt ~\sum_{b,a}\gamma_{ba}\rho_{aa}(t)=a_i(\epsilon_i)~.
\eeq
The rate of the transition to $b$-state given $a$-state for $a\neq b$ is
\beq
\langle b|~({\cal L} |a\rl a|~|b\rangle=\langle b |\sum_{b',a'}L_{b'a'}|a\rl a| L_{b'a'}^\dagger| b\rangle=\gamma_{ba}~.
\eeq
$a(\epsilon)$, called dynamical activity, is the expectation value of the number of transitions occurring during the time interval $\tau$ given a projected state $|\epsilon_i\rangle$. 
The total dynamical activity is given as 
\beq
{\cal A}=\sum_{i=1}^{N-1}\sum_{\ei}a_i(\ei) P_i(\ei)
\label{dynamical_activity}
\eeq

For R2, we extend the derivation by Hasegawa to time-dependent state. Eq.~(\ref{QFI_detail}) gives
\beqn
{\cal QF}_i&=&\int_0^\tau \!\!dt \sum_{b,a} \!\!\left[\sqrt{\gamma_{ba}\rho_{aa}(t)}-\sqrt{\gamma_{ab}\rho_{bb}(t)}\right]^2
\nonumber\\
&\le& \frac{1}{2} \!\int_0^\tau\!\! dt \sum_{b,a}\left[\gamma_{ba}\rho_{aa}(t)-\gamma_{ab}\rho_{bb}(t)\right]
\ln\!\sqrt{\frac{\gamma_{ba}\rho_{aa}(t)}{\gamma_{ab}\rho_{bb}(t)} }
\nonumber\\ 
&=&\frac{1}{2}\int_0^\tau\!\! dt \sum_{b,a}\gamma_{ba}\rho_{aa}(t)\ln\!\left[\frac{\gamma_{ba}\rho_{aa}(t)}{\gamma_{ab}\rho_{bb}(t)} \right]
\nonumber\\
&=& \frac{1}{2}\sigma_i(\epsilon_i)
\label{EP}
\eeqn
where Hasegawa showed the inequality using $(x-y)^2\le \frac{1}{2} (x^2-y^2)\ln\frac{x}{y}~(x, y>0)$. Let us consider the rate of dimensionless entropy change:
$\dot{E}=\textrm{Tr}\left(-\ln\rho-\beta H\right){\cal L}\rho$ where the first term is the rate of change in the von Neumann entropy of the system and the second is the rate of heat extraction divided by temperature to the reservoir. Using Eq.~(\ref{deriv_generator}), we find
\beqn
\dot{E}&=&\textrm{Tr}\sum_{b,a}\gamma_{ba}\Big[|a\rl a|(\ln{\rho}_{aa}+\beta\epsilon_a)\rho_{aa}
+|b\rl b|\rho_{aa}(-\ln\rho_{bb}-\beta\epsilon_{b}) \Big]
\nonumber\\
&=&\sum_{b,a}\gamma_{ba}\rho_{aa}\left[\ln\rho_{aa}-\ln\rho_{bb}+\beta(\epsilon_a-\epsilon_b)\right]~.
\eeqn
If we use the detailed balance condition $\gamma_{ab}=e^{-\beta(\epsilon_a-\epsilon_b)}\gamma_{ba}$, $\sigma_i(\ei)$ in Eq.~(\ref{EP}) turns out to be the entropy production $\int_0^\tau dt \dot{E}$ for the $i$-th interval given an initial $\ei$-state. From Eq.~(\ref{EP}), note that it is not the QFI in the first line, but the upper bound of it. The total dimensionless entropy production over all intervals is given as
\beq
{\cal E}=\sum_{i=1}^{N-1}\sum_{\ei}\sigma_i(\ei) P_i(\ei)~.
\eeq

For completeness, we also examined the perturbation rule R3 considered by Vu~\cite{vu2025}.  For each switching interval it yields the local bound
\begin{equation}
{\cal QF}^{(R3)}_i\le
\min\left[\frac{\sigma_i(\epsilon_i)}{2},a_i(\epsilon_i)\right].
\end{equation}
For a time-dependent protocol, however, different intervals may be limited by different branches of this minimum.  Consequently, summing the local bounds does not in general produce a global QFI bounded by $\min({\cal E}/2,{\cal A})$.  R3 is therefore not required for the global QTUR; the independent R1 and R2 bounds are sufficient.

Finally, we find the QTUR as
\beq
\frac{var W}{\left[\tau\partial_\tau\langle W\rangle\right]^2}\ge \frac{1}{\cal F}\ge \textrm{max}\left( \frac{1}{\cal A},\frac{2}{\cal E}\right)~.
\eeq
This global bound combines two independent perturbations of the same FI: R1 gives ${\cal F}\le {\cal QF}^{(R1)}={\cal A}$, while R2 gives ${\cal F}\le {\cal QF}^{(R2)}\le {\cal E}/2$. 

\section{A two-level system}

We use the Pauli matrices: $\sigma_1,\sigma_2,\sigma_3$ and $\sigma_{\pm}=(\sigma_1\pm i\sigma_2)/2$.
In $H_i$ basis, the matrix representation gives
 \beq
 H_i=(\hbar\omega_i/2)\sigma_3~, L_{i}^{\mp}=\sqrt{\gamma_{i}^{\mp}}\,\sigma_{\mp}
 \eeq
 where $\gamma_{i}^{-}=\gamma_0 (1+N(\omega_i))$, $\gamma_{i}^{+}=\gamma_0 N(\omega_i)$, and $\omega_i=\omega_0\sqrt{1+\alpha_i^2}$.
 For convenience, we vectorize density operator
\beq
\rho\to \widehat{\rho}=\sum_{m,n}\rho_{mn}| m\rangle \otimes | n\rangle~.
\eeq
Then, superoperators are transformed as
\beq
A\rho\to A\otimes I~ \widehat{\rho},~\rho A\to I\otimes A^{\textrm{T}}\widehat{\rho}
\eeq 
where $I$ is the $2 \times 2$ identity matrix and the superscript $\textrm{T}$ denotes transpose. Then, we have
\beqn
\widehat{{\cal L}}_i&=&-i\frac{\omega_i}{2}(\sigma_3\otimes I-I\otimes \sigma_3)
-\frac{\gamma_{i}^{-}}{2}\left[\left(\sigma_{+}\sigma_{-}\right)\otimes I+I\otimes \left(\sigma_{+}\sigma_{-}\right)\right]
-\frac{\gamma_{i}^{+}}{2}\left[\left(\sigma_{-}\sigma_{+}\right)\otimes I+I\otimes \left(\sigma_{-}\sigma_{+}\right)\right]
\nonumber\\
&&+\gamma_{i}^{-}\sigma_{-}\otimes \sigma_{+}+\gamma_{i}^{+}\sigma_{+}\otimes \sigma_{-}
\label{widehat_L}
\eeqn
where we use $A\otimes B C\otimes D=AC\otimes BD$. 
We introduce the dimensionless quantities 
\beq
\bar{t}=\omega_0 t,~\bar{\gamma}_0=\gamma_0/\omega_0,~\bar{\beta}=\beta\hbar\omega_0,~\bar{{\cal L}}={\cal L}/\omega_0~.
\eeq
The corresponding dimensionless vectorized generator is
\beq
\widehat{\bar{{\cal L}}}_i=\left(\begin{array}{cccc}
-\bar{\gamma}_i^{-} &0&0& \bar{\gamma}_i^{+}\\
0&-\frac{\bar{\gamma}_i}{2}+ix_i&0&0\\
0&0&-\frac{\bar{\gamma}_i}{2}-ix_i\\
\bar{\gamma}_i^{-}&0&0&-\bar{\gamma}_i^{+} 
\end{array}
\right)
\eeq
where $\bar{\gamma}_i^{-}=\bar{\gamma}_0(1+N(\omega_i))$, $\bar{\gamma}_i^{+}=\bar{\gamma}_0 N(\omega_i)$, $\bar{\gamma}_i=\bar{\gamma}_i^{+}+\bar{\gamma}_i^{-}$, and $x_i=\sqrt{1+\alpha_i^2}$. More explicitly, we have 
\beq
\bar{\gamma}_i=\bar{\gamma}_0\coth\left(\bar{\beta}x_i/2\right)~,~
\bar{\gamma}_i^{-}=\bar{\gamma}_0\coth\left(\bar{\beta}x_i/2\right)\frac{e^{\bar{\beta}x_i}}{e^{\bar{\beta}x_i}+1}~,~\bar{\gamma}_i^{+}=\bar{\gamma}_0\coth\left(\bar{\beta}x_i/2\right)\frac{1}{e^{\bar{\beta}x_i}+1}
\eeq

The dimensionless time-evolution operator is $\widehat{T}_i=e^{\bar{\tau}\widehat{\bar{{\cal L}}}_i}=\lim_{n\to \infty}(1+d\bar{t} ~\widehat{\bar{{\cal L}}}_i)^n$. 
A direct calculation gives
\beqn
(I+d\bar{t}~\widehat{\bar{{\cal L}}}_i)^n
&=&\left(
\begin{array}{cccc}
c^n+d\bar{t}\bar{\gamma}_i^{+}\sum_{m=0}^{n-1}c^m&0&0&d\bar{t}\bar{\gamma}_i^{+}\sum_{m=0}^{n-1}c^m\\
0&\left[1-d\bar{t}(\bar{\gamma}_i/2-ix_i)\right]^n&0&0\\
0&0&\left[1-d\bar{t}(\bar{\gamma}_i/2+ix_i)\right]^n&0\\
d\bar{t}\bar{\gamma}_i^{-}\sum_{m=0}^{n-1}c^m&0&0&
c^n+d\bar{t}\bar{\gamma}_i^{-}\sum_{m=0}^{n-1}c^m
\end{array}
\right)
\eeqn
where $c=1-d\bar{t}~ \bar{\gamma}_i$. In the limit $n\to\infty$, we find 
\beqn
e^{\bar{\tau}\widehat{\bar{{\cal L}}}_i}
&=&\left(\begin{array}{cccc}
e^{-\bar{\tau}\bar{\gamma}_i}+\frac{\bar{\gamma}_i^{+}}{\bar{\gamma}_i}(1-e^{-\bar{\tau}\bar{\gamma}_i})&0&0&\frac{\bar{\gamma}_i^{+}}{\bar{\gamma}_i}(1-e^{-\bar{\tau}\bar{\gamma}_i})\\
0&e^{-\bar{\tau}(\bar{\gamma}_i/2-ix_i)}&0&0\\
0&0&e^{-\bar{\tau}(\bar{\gamma}_i/2+ix_i)}&0\\
\frac{\bar{\gamma}_i^{-}}{\bar{\gamma}_i}(1-e^{-\bar{\tau}\bar{\gamma}_i})&0&0&
e^{-\bar{\tau}\bar{\gamma}_i}+\frac{\bar{\gamma}_i^{-}}{\bar{\gamma}_i}(1-e^{-\bar{\tau}\bar{\gamma}_i})
\end{array}
\right)
\label{time_evolution}
\eeqn

We can write the generating function for the time evolution through two-point measurements from repeated matrix products shown in Eq.~(2) in the main text: 
\beq
{\cal G}_N(\lambda)=(1,1)\cdot G_N(\lambda) \cdot T_{N-1}\cdot P_{N-1}(\lambda)
\eeq
where the recursion relation is given by
\beq
P_i(\lambda)=G_{i}(\lambda)\cdot T_{i-1}\cdot P_{i-1}(\lambda)~~\textrm{for $2\le i\le N-1$}
\label{recursion_relation}
\eeq
and
\beq
P_1(\lambda)=\frac{1}{e^{-\bar{\beta}/2}+ e^{\bar{\beta}/2}}
\left(\begin{array}{c}
e^{-\bar{\beta}/2}\\ e^{\bar{\beta}/2}
\end{array}
\right)~,
\eeq
that does not depend on $\lambda$. 

$G_i$ and $T_i$ are $2\times 2$ matrices, and $P_i$ is a 2-component column vector:
\beqn
[G_i]_{\epsilon_{i+1},\eip}&=&e^{-\beta\lambda(\epsilon_{i+1}-\eip)}|\langle\epsilon_{i+1}| \eip\rangle |^2~,
\nonumber\\
{[T_i]}_{\eip,\ei}&=&\langle \eip |  \left[e^{\tau{\cal L}_i} |\ei \rl \ei |\right] |\eip\rangle~.
\eeqn
For the two-level case,
\beqn
| \epsilon_i^{+}\rangle&=&N_i^{-1}\left(\left(1+\sqrt{1+\alpha_i^2}\right) |e\rangle+\alpha_i |g\rangle\right),~
\nonumber\\
| \epsilon_i^{-}\rangle&=&N_i^{-1}\left(-\alpha_i |e\rangle+\left(1+\sqrt{1+\alpha_i^2}\right)|g \rangle\right),~
\nonumber\\
\epsilon_i^{\pm}&=&\pm \hbar\omega_0\sqrt{1+\alpha_i^2}/2
\eeqn
where $N_i=\left[\alpha_i^2+\left(1+\sqrt{1+\alpha_i^2}\right)^2\right]^{1/2}$. Using this equation, $G_i$ can be found. $|\ei^{\pm})$ is defined as the vectorization of $|\ei^\pm  \rl \ei^\pm |$. In $H_i$-basis, $|\ei^{+})=(1,0,0,0)^{\textrm{T}}$ and $|\ei^{-})=(0,0,1,0)^{\textrm{T}}$. Therefore, we can find
\beq
{[T_i]}_{\eip,\ei}=({\ei}^{\pm}|e^{\bar{\tau}\widehat{\bar{{\cal L}}}_i} |\ei^{\pm})=\widehat{\rho}_i^{\pm\pm}(\tau)=\left(\begin{array}{cc}
\widehat{\rho}_i^{++}(\tau)&\widehat{\rho}_i^{+-}(\tau)\\
\widehat{\rho}_i^{-+}(\tau)&\widehat{\rho}_i^{--}(\tau)
\end{array}
\right)
\label{T_comp}
\eeq
for all combinations of $\pm$ signs. Using Eq.~(\ref{time_evolution}), 
\beq
T_i=\left(\begin{array}{cc}
e^{-\bar{\tau}\bar{\gamma}_i}+\frac{\bar{\gamma}_i^{+}}{\bar{\gamma}_i}(1-e^{-\bar{\tau}\bar{\gamma}_i})&\frac{\bar{\gamma}_i^{+}}{\bar{\gamma}_i}(1-e^{-\bar{\tau}\bar{\gamma}_i})\\
\frac{\bar{\gamma}_i^{-}}{\bar{\gamma}_i}(1-e^{-\bar{\tau}\bar{\gamma}_i})&
e^{-\bar{\tau}\bar{\gamma}_i}+\frac{\bar{\gamma}_i^{-}}{\bar{\gamma}_i}(1-e^{-\bar{\tau}\bar{\gamma}_i})
\end{array}\right)
\label{rho_comp}
\eeq

We find expressions necessary for the QTUR in terms of matrices and their derivatives with respect to $\lambda$. For simple notations, $f^{(n)}=(\partial_\lambda^n f(\lambda)|_{\lambda\to 0}$ for $n=0,1,2$.
The expectation value of work is given by
\beq
\langle{\cal W}\rangle=\sum_i^N (1,1)\cdot G_i^{(1)}\cdot T_i\cdot P_{i-1}^{(0)}~.
\eeq
The variance of work is given by $\langle {\cal W}^2\rangle -\langle{\cal W}\rangle^2$ where
\beq
\langle {\cal W}^2\rangle=(1,1)\!\cdot\!\left[ G_N^{(2)}\!\!\cdot\! T_{N-1}\!\cdot\! P_{N-1}^{(0)}+2 G_N^{(1)}\! \!\cdot\! T_{N-1}\!\!\cdot\! P_{N-1}^{(1)}\!+\!G_N\! (0)\!\cdot \!T_{N-1}\!\!\cdot\! P_{N-1}^{(2)}\!\right]~.
\eeq 
where recursion relations are:
\beqn
P_{i}^{(0)}&=&G_i^{(0)}\!\!\cdot\! T_{i-1}\!\cdot\! P_{i-1}^{(0)}
~,\nonumber\\
P_{i}^{(1)}&=&G_i^{(1)}\!\!\cdot\! T_{i-1}\!\cdot\! P_{i-1}^{(0)}+G_i^{(0)}\!\!\cdot\! T_{i-1}\!\cdot\! P_{i-1}^{(1)}
~,\nonumber\\
P_{i}^{(2)}&=&G_i^{(2)}\!\!\cdot\! T_{i-1}\!\cdot\! P_{i-1}^{(0)}
+2G_i^{(1)}\!\!\cdot\! T_{i-1}\!\cdot\! P_{i-1}^{(1)}
+G_i^{(0)}\!\!\cdot\! T_{i-1}\!\cdot\! P_{i-1}^{(2)}~.
\eeqn
From Eq.~(\ref{work_deriv}), 
\beq
\p\langle W\rangle = \tau\sum_i^N (1,1)\cdot G_i^{(1)}\cdot \left[\partial_\tau T_{i-1} \cdot 
P_{i-1}^{(0)}
 +T_{i-1}\cdot \partial_\tau P_{i-1}^{(0)}\right]~,
\eeq
where the necessary recursion relation is
\beq
\partial_\tau P_{i-1}^{(0)}=G_{i-1}^{(0)}\cdot\left[ \partial_\tau T_{i-2}\cdot P_{i-2}^{(0)}
+T_{i-2}\cdot \partial_\tau P_{i-2}^{(0)}\right]~.
\eeq
 
From Eq.~(\ref{FI_tau}), the FI is given
\beq
{\cal F}=\sum_{i=1}^{N-1}(1,1)\cdot F_i\cdot P_i^{(0)}
\eeq
where the elements of the matrix $F_i$ are given using Eq.~(\ref{rho_comp})
\beq
[{F_i}]_{\epsilon_i^{\pm},\epsilon_i^{\pm}}=\frac{\left[ \tau\partial_\tau [T_i]_{ \epsilon_i^{\pm},\epsilon_i^\pm}\right]^2}{[T_i]_{ \epsilon_i^{\pm},\epsilon_i^\pm}}
\eeq 
for all combinations of $\pm$ signs. 

Let us define:
\beq
\widehat{\rho}_i^{\pm,\pm}(\bar{t})=(\epsilon_i^{\pm}| e^{\bar{t}\widehat{\bar{\cal L}}_i} |\epsilon_i^{\pm})~,
\eeq
which is equal to $T_i$ in Eq.~(\ref{T_comp}) for $\bar{t}=\bar{\tau}$. 
The dynamical activity is found from Eq.~(\ref{dynamical_activity})
\beq
{\cal A}=\sum_{i=1}^{N-1} \int_0^{\bar{\tau}}\!d\bar{t}~(1,1)\cdot A_i(\bar{t})\cdot P_i^{(0)}~.
\label{activity_integral}
\eeq
The matrix $A_i(t)$ is given by 
\beq
A_i(\bar{t})=\left(\begin{array}{cc}\bar{\gamma}_i^{-}\widehat{\rho}_i^{++}(\bar{t})& \bar{\gamma}_i^{-}\widehat{\rho}_i^{+-}(\bar{t})\\
\bar{\gamma}_i^{+}\widehat{\rho}_i^{-+}(\bar{t})& \bar{\gamma}_i^{+}\widehat{\rho}_i^{--}(\bar{t})
\end{array}
\right)~,
\eeq
where matrix elements can be found from Eq.~(\ref{rho_comp}) by replacing $\bar{\tau}$ by $\bar{t}$.
The entropy production is given
\beq
{\cal E}=\sum_{i=1}^{N-1} \int_0^{\bar{\tau}}\!d\bar{t}~(1,1)\cdot E_i(\bar{t})\cdot P_i(0)~,
\label{EP_integral}
\eeq
where the matrix $E_i(\bar{t})$ is given 
\beq
E_i(\bar{t})=\left(\begin{array}{cc}\bar{\gamma}_i^{-}\widehat{\rho}_i^{++}(\bar{t})R_i^{++}(\bar{t})& \bar{\gamma}_i^{-}\widehat{\rho}_i^{+-}(\bar{t})R_i^{+-}(\bar{t})\\
\bar{\gamma}_i^{+}\widehat{\rho}_i^{-+}(\bar{t})R_i^{-+}(\bar{t})& \bar{\gamma}_i^{+}\widehat{\rho}_i^{--}(\bar{t})R_i^{--}(\bar{t})
\end{array}
\right)~,
\eeq
and
\beqn
R_i^{+\pm}(\bar{t})&\!=\!&\ln\!\frac{\bar{\gamma}_i^{-}\widehat{\rho}_i^{+\pm}(\bar{t})}{\bar{\gamma}_i^{+}\widehat{\rho}_i^{-\pm}(\bar{t})}~,\\
R_i^{-\pm}&=&-R_i^{+\pm}~.
\eeqn
The integrals in Eqs.~(\ref{activity_integral}) and (\ref{EP_integral}) can be evaluated analytically; the resulting expressions are not shown here. We use the analytical expressions for matrices in this section and execute the matrix products repeatedly from $i=1$ to $N$ numerically for numerical parameters such as $\bar\beta=1,\bar{\gamma}_0=0.1, \bar{\tau}=1, 3,5,7$, etc.

\end{document}